\headline={\ifnum\pageno>1 \hss \number\pageno\ \hss \else\hfill \fi}
\pageno=1
\nopagenumbers

\centerline{\bf NON-RECURSIVE MULTIPLICITY FORMULAS FOR $A_N$ LIE ALGEBRAS}
\vskip 15mm
\centerline{\bf H. R. Karadayi}
\centerline{Dept.Physics, Fac. Science, Tech.Univ.Istanbul }
\centerline{ 80626, Maslak, Istanbul, Turkey }
\centerline{ Internet: karadayi@sariyer.cc.itu.edu.tr}
\vskip 10mm
\centerline{\bf{Abstract}}
\vskip 10mm

It is shown that there are infinitely many formulas to calculate multiplicities
of weights participating in irreducible representations of $A_N$ Lie
algebras. On contrary to the recursive character of Kostant and Freudenthal
multiplicity formulas, they provide us systems of linear algebraic equations
with N-dependent polinomial coefficients. These polinomial coefficients are
in fact related with polinomials which represent eigenvalues of Casimir
operators.

\vskip 15mm
\vskip 15mm
\vskip 15mm
\vskip 15mm
\vskip 15mm
\vskip 15mm
\vskip 15mm
\vskip 15mm

\hfill\eject

\vskip 3mm
\noindent {\bf{I.\ INTRODUCTION}}
\vskip 3mm
In a previous paper {\bf [1]}, we establish a general method to calculate
eigenvalues of Casimir operators of any order. Although it is worked for
$A_N$ Lie algebras in ref.(1), the method works well also for other
Classical and Exceptional Lie algebras because any one of them has a subgroup
of type $A_N$. Let I(s) be a Casimir operator of degree s and
$$ s_1 + s_2 + ... + s_k \equiv s  \eqno(I.1)  $$
be the partitions of s into positive integers on condition that
$$ s_1 \geq s_2 \geq ... \geq s_k  \ \ . \eqno(I.2) $$
The number of all these partitions of s is given by the partition
function p(s) which is known to be defined by
$$ \prod_{n=1}^\infty {1 \over (1-x^n)} \equiv \sum_{s=0}^\infty p(s) \ x^s \ \ . $$
We also define $\kappa(s)$ to be the number of partitions of s into positive
integers except 1. We know then that we have at least a $ \kappa(s) $
number of ways to represent the eigenvalues of I(s) in the form of polinomials
$$ P_{s_1,s_2, .. s_k}(\Lambda^+,N) \eqno(I.3) \ \ .  $$
where $\Lambda^+$ is the dominant weight which specifies the representation
$R(\Lambda^+)$ for which we calculate eigenvalues. In ref.(1), we give,
for orders s=4,5,6,7  all the polinomials (I.3) explicitly and show that they
are in coincidence with the ones calculated directly as traces of the most
general Casimir operators {\bf [2]}. This, on the other hand, brings out
another problem of computing multiplicities of all other weights participating
within the representation $ R(\Lambda^+) $. This problem has been solved some
forty years ago by Kostant {\bf [3]} and also Freudenthal {\bf[4]}. In spite of
the fact that they are quite explicit ,the multiplicity formulas of Kostant
and Freudenthal expose serious difficulties in practical calculations due
especially to their recursive characters. It is therefore worthwhile to try
for obtaining some other non-recursive formulas to calculate weight
multiplicities. Let us recall here that,since the last twenty years
{\bf [5]}, there are efforts which give us weight multiplicities in the
form of tables, one by one.

We will  give in this work general formulas with the property that they
depend only on multiplicities and rank N. This will give us the possibility
to obtain, by induction on values N=1,2,... , as many equations as we need
to solve all unknown parameters, i.e. multiplicities.

\vskip 3mm
\noindent {\bf{II.\ WEYL ORBITS AND IRREDUCIBLE REPRESENTATIONS }}
\vskip 3mm
We refer the book of Humphreys {\bf [6]} for Lie algebraic technology which
we need in the sequel. The indices like $i_1,i_2,..$ take values from the
set $I_\circ \equiv {1,2, .. N}$ while $I_1,I_2,..$ from $
S_\circ \equiv {1,2, .. N,N+1}$. The essential figures are simple roots
$\alpha_i$, fundamental dominant weights $\lambda_i$ and the fundamental
weights defined by
$$ \eqalign{
\mu_1 &\equiv \lambda_1 \cr
\mu_I &\equiv \mu_{I-1} - \alpha_{I-1} \ \ , \ \ I=2,3,.. N+1 } \ \ .\eqno(II.1) $$
Any dominant weight then has the decompositions
$$ \Lambda^+ = \sum_{i=1}^N r_i \ \lambda_i \ \ , \ \ r_i \in Z^+ \eqno(II.2) $$
where $Z^+$ is the set of positive integers including zero.
(II.2) can be expressed equivalently in the form
$$ \Lambda^+ = \sum_{i=1}^\sigma q_i \ \mu_i \ \ , \ \
\sigma = 1,2, .. N  \eqno(II.3) $$
together with the conditions
$$ q_1 \geq q_2 \geq ... \geq q_\sigma > 0  \ \ . \eqno(II.4) $$
The weight space decomposition of an irreducible representation $R(\Lambda^+)$
has now the form
$$ R(\Lambda^+) = \Pi(\Lambda^+) \ \ + \sum_{\lambda^+ \in Sub(\Lambda+^+)}
m(\lambda^+ < \Lambda^+) \ \ \Pi(\lambda^+) \eqno(II.5) $$
where $m(\lambda^+ < \Lambda^+)$ is the multiplicity of weight $\lambda^+$
within the representation $R(\Lambda^+)$ and $\Pi(\lambda^+)$ represents
its corresponding Weyl orbit. $Sub(\Lambda^+)$ here is the set of all
sub-dominant weights of $\Lambda^+$. Although the concept of Casimir
eigenvalue is known to be defined for representations of Lie algebras,
an extension for Weyl orbits has been made, in ref.(1), with the aid of
definition
$$ ch_s(\Pi) \equiv \sum_{\mu \in \Pi} (\mu)^s \ \ . \eqno(II.6) $$
This allows us to make the decompositions
$$ ch_s(\Pi) = \sum_{s_1,s_2, .. s_k} \mu(s_1) \ \mu(s_2) ... \mu(s_k) \ \
cof_{s_1,s_2, .. s_k}  \eqno(II.7) $$
in terms of generators $ \mu(s) $ defined by
$$ \mu(s) \equiv \sum_{I=1}^{N+1} (\mu_I)^s \ \ . \eqno(II.8) $$
We consider here the partitions (I.1) for any order s. It is then known that
the coefficients in (II.7) are expressed in the form
$$ cof_{s_1,s_2, .. s_k} \equiv cof_{s_1,s_2, .. s_k}(\lambda^+,N) \ \ , $$
i.e. as N-dependent polinomials for a Weyl orbit $ \Pi(\lambda^+) $ of $ A_N $
Lie algebras. One further step is to propose the existence of some polinomials
$ P_{s_1,s_2, .. s_k}(\lambda^+,N) $ satisfying following equations:
$$ P_{s_1 s_2 .. s_k}(\lambda^+,N) \equiv {cof_{s_1 s_2 .. s_k}(\lambda^+,N) \over
cof_{s_1 s_2 .. s_k}(\lambda_k,N)} \ {dimR(\lambda_k,N) \over dimR(\lambda^+,N)} \
P_{s_1 s_2 .. s_k}(\lambda_k,N) \ \ . \eqno(II.9) $$
Note here that
$$ cof_{s_1 s_2 .. s_k}(\lambda_i,N) \equiv 0 \ \ \ , \ \ \ i < k  \eqno(II.10) $$
and also
$$ dimR(\lambda_k,N) = { (N+1)! \over k! \ (N+1-k)! } \ \ \ , \ \ \
k=1,2, .. N. \eqno(II.11)  $$
where $ dimR(\Lambda^+) $ is the number of weights within the representation
$R(\Lambda^+)$, i.e. its dimension. By examining (II.9) for a few simple
representations, a $ \kappa(s) $ number of polinomials can be obtained for each
particular value of order s. The ones for s=4,5,6,7 are given in ref.(1)
explicitly.

\vskip 3mm
\noindent {\bf{III.\ THE MULTIPLICITY FORMULAS}}
\vskip 3mm
Our strategy here is to use the equations (II.9) in the form of

$$ \eqalign{
\Phi^{s_1,s_2, .. s_k}(\lambda^+,N) \equiv
&P_{s_1 s_2 .. s_k}(\lambda^+,N) \ dimR(\lambda^+,N) \
cof_{s_1 s_2 .. s_k}(\lambda_k,N) \ - \cr
&P_{s_1 s_2 .. s_k}(\lambda_k,N) \ dimR(\lambda_k,N) \
cof_{s_1 s_2 .. s_k}(\lambda^+,N) }  \eqno(III.1) $$
with which we obtain, for each particular partition (I.1) of degree s,
a multiplicity formula
$$ \Phi^{s_1,s_2, .. s_k}(\lambda^+,N) \equiv 0 \eqno(III.2)   $$
for weight multiplicities within the representation $R(\lambda^+)$ and
for any values of rank $ N > \sigma $. We think $\lambda^+$ here as in the
form of (II.3). In addition to the ones given in (II.10), the following
expressions can be borrowed from ref.(1):
$$ \eqalign{
&cof_s(\lambda_1,n)=1 \ \ , \cr
&cof_{s_1,s_2}(\lambda_2,n)={ (s_1+s_2)! \over s_1! \ s_2!} \ \ , \ \ s_1 > s_2 \ \ , \cr
&cof_{s_1,s_1}(\lambda_2,n)={ (s_1+s_1)! \over 2! \ s_1! \ s_1!} \ \ , \cr
&cof_{s_1,s_2,s_3}(\lambda_3,n)={ (s_1+s_2+s_3)! \over s_1! \ s_2! \ s_3!} \ \ , \ \ s_1 > s_2 > s_3 \ \ , \cr
&cof_{s_1,s_2,s_2}(\lambda_3,n)={ (s_1+s_2+s_2)! \over 2! \ s_1! \ s_2! \ s_2!} \ \ , \ \ s_1 > s_2  \ \ , \cr
&cof_{s_1,s_1,s_1}(\lambda_3,n)={ (s_1+s_2+s_1)! \over 3! \ s_1! \ s_1! \ s_1!} \ \ . } \eqno(III.3) $$
These are sufficient to give
$$ \sum_{s=4}^7 \kappa(s) = 12 $$
different multiplicity formulas originating from the same form (III.2).
To proceed further let us take (II.5) in the form
$$ R(\Lambda^+) \equiv \sum_{\alpha=1}^{p(h_{\Lambda^+})} m(\alpha) \
\Pi(\rho_\alpha) \ \ , \ \  \rho_\alpha \in Sub(\Lambda+^+)  \eqno(III.4) $$
where we usually define the {\bf height}
$$ h_{\Lambda^+} \equiv \sum_{i=1}^\sigma q_i \eqno(III.5) \ \ . $$
for (II.3) and $ p(h_{\Lambda^+}) $ is just the partition function defined
above. A further focus here is to make a gradation for elements of the set
$Sub(s \ \lambda_1)$ by assigning, for each one of them, a grade
$$ \gamma(s_1,s_2, .. ,s_k) = 1,2, .. ,p(s)  \eqno(III.6) $$
as being in line with the conditions (I.2).
Then, it is clear in (III.4) that
$$ m(\alpha) \equiv 0 \ \ \ , \ \ \ \gamma(\rho_\alpha) >
\gamma(\Lambda^+) \ \ . \eqno(III.7) $$
Note also that all dominant weights within a $Sub(\Lambda^+)$ must have the
same height.

In view of (III.4), one knows both $ cof_{s_1 s_2 .. s_k}(\lambda^+,N) $ and
also $ dimR(\lambda^+,N) $ as linear superpositions of multiplicities
$ m(\alpha) $ with N-dependent polinomial coefficients. It is noteworthy
here that dimensions of Weyl orbits are already known due to a permutational
lemma given in ref.(1). We ,hence, give in the following our results for 12
multiplicity formulas extracted from (III.2) for s=4,5,6,7:
$$ \eqalign{ \Phi^7(\lambda^+,N)=
cof_7(\lambda^+,N) \ g(N) \ &+  \cr
dimR(\lambda^+,N) \ \ ( \ \
 - \   720  \ f_7^7(N)     \ &\Theta(7,\lambda^+,N)                                 \cr
 + \  5040  \ f^7_{52}(N)  \ &\Theta(5,\lambda^+,N) \ \Theta(2,\lambda^+,N)         \cr
 + \  5040  \ f^7_{43}(N)  \ &\Theta(4,\lambda^+,N) \ \Theta(3,\lambda^+,N)         \cr
 - \ 10080  \ f^7_{322}(N) \ &\Theta(3,\lambda^+,N) \ \Theta(2,\lambda^+,N)^2 \ )
} \eqno(III.8)   $$

$$ \eqalign{ \Phi^{52}(\lambda^+,N)=
cof_{52}(\lambda^+,N) \ N \ (N + 2) \ g(N) \ &+ \cr
dimR(\lambda^+,N) \ \ ( \ \
     5040 \ f^{52}_7(N)     \ &\Theta(7,\lambda^+,N)                                \cr
 - \  504 \ f^{52}_{52}(N)  \ &\Theta(5,\lambda^+,N)  \ \Theta(2,\lambda^+,N)       \cr
 - \ 5040 \ f^{52}_{43}(N)  \ &\Theta(4,\lambda^+,N)  \ \Theta(3,\lambda^+,N)       \cr
 + \ 2520 \ f^{52}_{322}(N) \ &\Theta(3,\lambda^+,N)  \ \Theta(2,\lambda^+,N)^2     \cr
 + \   42 \ f^{52}_5(N)     \ &\Theta(5,\lambda^+,N)                                \cr
 - \  210 \ f^{52}_{32}(N)  \ &\Theta(3,\lambda^+,N)  \ \Theta(2,\lambda^+,N) \ )
} \eqno(III.9)   $$

$$ \eqalign{ \Phi^{43}(\lambda^+,N)=
cof_{43}(\lambda^+,N) \ 12 \ N \ (N + 2) \ g(N) \ &+ \cr
dimR(\lambda^+,N) \ \ ( \ \
     60480 \ f^{43}_7(N)     \ &\Theta(7,\lambda^+,N)                             \cr
 - \ 60480 \ f^{43}_{52}(N)  \ &\Theta(5,\lambda^+,N) \ \Theta(2,\lambda^+,N)     \cr
 - \ 5040  \ f^{43}_{43}(N)  \ &\Theta(4,\lambda^+,N) \ \Theta(3,\lambda^+,N)     \cr
 + \ 5040  \ f^{43}_{322}(N) \ &\Theta(3,\lambda^+,N) \ \Theta(2,\lambda^+,N)^2   \cr
 - \    7  \ f^{43}_3(N)     \ &\Theta(3,\lambda^+,N) \ )
} \eqno(III.10)   $$

$$ \eqalign{ \Phi^{322}(\lambda^+,N)=
cof_{322}(\lambda+,N) \ 24 \ N \ (N + 2) \ g(N) \ &+ \cr
dimR(\lambda^+,N) \ \ ( \ \
 - \ 241920  \ f^{322}_7(N)     \ &\Theta(7,\lambda^+,N)                               \cr
 + \  60480  \ f^{322}_{52}(N)  \ &\Theta(5,\lambda^+,N) \ \Theta(2,\lambda^+,N)    \cr
 + \  10080  \ f^{322}_{43}(N)  \ &\Theta(4,\lambda^+,N) \ \Theta(3,\lambda^+,N)    \cr
 - \   5040  \ f^{322}_{322}(N) \ &\Theta(3,\lambda^+,N) \ \Theta(2,\lambda^+,N)^2   \cr
 - \   5040  \ f^{322}_5(N)     \ &\Theta(5,\lambda^+,N)                            \cr
 - \    840  \ f^{322}_{32}(N)  \ &\Theta(3,\lambda^+,N) \ \Theta(2,\lambda^+,N)     \cr
 - \      7  \ f^{322}_3(N)     \ &\Theta(3,\lambda^+,N) \ )
} \eqno(III.11)   $$

$$ \eqalign{ \Phi^6(\lambda^+,N)=
cof_6(\lambda^+,N) \ 252 \ g_6(N) \ +
dimR(\lambda^+, N) \ (N + 1) \ g(N) \ &+                                        \cr
dimR(\lambda^+,N) \ \ ( \ \
- \  30240 \ f^6_6(N)     \  &\Theta(6,\lambda^+,N)                             \cr
+ \ 181440 \ f^6_{42}(N)  \  &\Theta(4,\lambda^+,N) \ \Theta(2,\lambda^+,N)     \cr
+ \  30240 \ f^6_{33}(N)  \  &\Theta(3,\lambda^+,N)^2                           \cr
- \ 211680 \ f^6_{222}(N) \  &\Theta(2,\lambda^+,N)^3 \ ) }  \eqno(III.12)  $$

$$ \eqalign{ \Phi^{42}(\lambda^+,N)=
cof_{42}(\lambda^+,N) \ 672 \ N \ (N + 1) \ (N + 2) \ g_6(N) \ &+\cr
dimR(\lambda^+, N) \ N \ (N + 1) \ (N + 2) \ (7 \ N^2 + & 14 \ N + 47) \ g_6(N)   \ + \cr
dimR(\lambda^+,N) \ \ ( \ \
  483840    \ f^{42}_6(N)       \ &\Theta(6,\lambda^+,N)                             \cr
- \ 60480   \ f^{42}_{42}(N)    \ &\Theta(4,\lambda^+,N) \ \Theta(2,\lambda^+,N)     \cr
- \ 1209600 \ f^{42}_{33}(N)    \ &\Theta(3,\lambda^+,N)^2                           \cr
+ \ 60480   \ f^{42}_{222}(N)   \ &\Theta(2,\lambda^+,N)^3                           \cr
+ \ 5040    \ f^{42}_4(N)       \ &\Theta(4,\lambda^+,N)                             \cr
- \ 5040    \ f^{42}_{22}(N)    \ &\Theta(2,\lambda^+,N)^2                           \cr
- \ 84      \ f^{42}_2(N)       \ &\Theta(2,\lambda^+,N) \ )  } \eqno(III.13)   $$

$$ \eqalign{ \Phi^{33}(\lambda^+,N)=
cof_{33}(\lambda^+,N) \ 126 \ N \ (N + 1) \ (N + 2) \ g_6(N) \ &-               \cr
dimR(\lambda^+,N) \ 5 \ N \ (N + 1) \ (N + 2) \ g_6(N) \ &+                     \cr
dimR(\lambda^+,N) \ \ ( \ \
     15120 \ f^{33}_6(N)     \  &\Theta(6,\lambda^+,N)                         \cr
- \ 226800 \ f^{33}_{42}(N)  \  &\Theta(4,\lambda^+,N) \ \Theta(2,\lambda^+,N) \cr
- \   5040 \ f^{33}_{33}(N)  \  &\Theta(3,\lambda^+,N)^2                       \cr
+ \  60480 \ f^{33}_{222}(N) \  &\Theta(2,\lambda^+,N)^3 \ )   } \eqno(III.14) $$

$$ \eqalign{ \Phi^{222}(\lambda^+,N)=
cof_{222}(\lambda^+,N) \ 576 \ N \ (N + 1) \ (N + 2) \ g_6(N) \ &+                   \cr
dimR(\lambda^+,N) \ N \ (N + 1)^2 \ (N + 2) \ &(5 \ N^2 + 10 \ N + 23) \ g_6(N) \  +  \cr
dimR(\lambda^+,N) \ \ ( \ \
    483840  \ f^{222}_6(N)     \ &\Theta(6,\lambda^+,N)                              \cr
- \  51840  \ f^{222}_{42}(N)  \ &\Theta(4,\lambda^+,N) \ \Theta(2,\lambda^+,N)      \cr
- \ 276480  \ f^{222}_{33}(N)  \ &\Theta(3,\lambda^+,N)^2                            \cr
+ \   8640  \ f^{222}_{222}(N) \ &\Theta(2,\lambda^+,N)^3                            \cr
+ \   4320  \ f^{222}_4(N)     \ &\Theta(4,\lambda^+,N)                              \cr
- \   2160  \ f^{222}_{22}(N)  \ &\Theta(2,\lambda^+,N)^2                            \cr
+ \     36  \ f^{222}_2(N)     \ &\Theta(2,\lambda^+,N) \ ) } \eqno(III.15)  $$

$$ \eqalign{ \Phi^5(\lambda^+,n_)=cof_5(\lambda^+,n) \ g_5(n) \ -
 \ dimR(\lambda^+,n) \ \ ( \ \ 24 \ f^5_5(n)  \ &\Theta(5,\lambda^+,n)    \cr
- \ 120 \ f^5_{32}(n) \                         &\Theta(3,\lambda^+,n) \ \Theta(2,\lambda^+,n) \  )  } \eqno(III.16)  $$

$$ \eqalign{ \Phi^{32}(\lambda^+,n_)=
cof_{32}(\lambda^+,n) \ 3 \ g_5(n) \ &+ \cr
\ dimR(\lambda^+,n) \ \ ( \ \
 360 \ f^{32}_5(n)    \ &\Theta(5,\lambda^+,n)                          \cr
- 60 \ f^{32}_{32}(n) \ &\Theta(3,\lambda^+,n) \ \Theta(2,\lambda^+,n)    \cr
+  5 \ f^{32}_3(n)    \ &\Theta(3,\lambda^+,n) \ ) } \eqno(III.17)   $$

$$ \eqalign{ \Phi^4(\lambda^+,N)=
- cof_4(\lambda^+,N) \ 120 \ g_4(N) \ + \
dimR(\lambda^+,N) \ (N + 1) \ g_4(N) \ &+  \cr
\ dimR(\lambda^+,N) \ \ ( \ \
    720 \ f^4_4(N)    \ &\Theta(4,\lambda^+,N)    \cr
- \ 720 \ f^4_{22}(N) \ &\Theta(2,\lambda^+,N)^2 \ )  } \eqno(III.18)   $$

$$ \eqalign{ \Phi^{22}(\lambda^+,N)=
cof_{22}(\lambda^+,N) \ 240 \ (N + 1) \ g_4(N) \ -
\ (N + 1) \ g22(N) dimR(\lambda^+,N) \ &+ \  \cr
dimR(\lambda^+,N) \ \ ( \ \
 1440 f224(N)  &\Theta(4,\lambda^+,N)      \cr
- 720 f2222(N) &\Theta(2,\lambda^+,N)^2    \cr
+ 120 f222(N)  &\Theta(2,\lambda^+,N) \  )  }  \eqno(III.19)   $$

\noindent where the quantities defined by
$$ \Theta(s_,\lambda^+,N) \equiv \sum_{I=1}^{N+1} (\theta_I)^s $$
can be calculated explicitly via re-definitions
$$ 1 + r_i \equiv \theta_i \ - \ \theta_{i+1} \ \ , \ \ i \in I_\circ   $$
of the parameters $r_i$ in (II.2). Note here that
$ \Theta(1,\lambda^+,N) \equiv 0 $. All coefficient polinomials are given in
appendix.

\hfill\eject

\vskip 3mm
\noindent {\bf{IV.\ AN EXAMPLE : $R(\lambda_1+\lambda_2+\lambda_6)$ }}
\vskip 3mm
Now it will be instructive to demonstrate the idea in an explicit example,
chosen, say, from the set $Sub(9 \ \lambda_1)$ with the gradation (III.6)
from 1 to p(9)=30 for its 30 elements all having the same height \ (= 9).
In the notation of parameters $ q_i $ defined in (II.3),
(III.4) turns out to be
$$ \eqalign{ R(\lambda_1+\lambda_2+\lambda_6) &=
   m(0) \ \Pi(3,2,1,1,1,1)               \cr
&+ m(1) \ \Pi(2,2,2,1,1,1)               \cr
&+ m(2) \ \Pi(3,1,1,1,1,1,1)             \cr
&+ m(3) \ \Pi(2,2,1,1,1,1,1)             \cr
&+ m(4) \ \Pi(2,1,1,1,1,1,1,1)           \cr
&+ m(5) \ \Pi(1,1,1,1,1,1,1,1,1) }           $$
with
$$ \eqalign{
dimR(\lambda_1+\lambda_2+\lambda_6) &=
   {m(0) \over 24} \ (N - 4) \ (N - 3) \ (N - 2) \ (N - 1) \ N \ (N + 1)                                       \cr
&+ {m(1) \over 36} \ (N - 4) \ (N - 3) \ (N - 2) \ (N - 1) \ N \ (N + 1)                                        \cr
&+ {m(2) \over 720} \ (N - 5) \ (N - 4) \ (N - 3) \ (N - 2) \ (N - 1) \ N \ (N + 1)                             \cr
&+ {m(3) \over 240} \ (N - 5) \ (N - 4) \ (N - 3) \ (N - 2) \ (N - 1) \ N \ (N + 1)                             \cr
&+ {m(4) \over 5040} \ (N - 6) \ (N - 5) \ (N - 4) \ (N - 3) \ (N - 2) \ (N - 1) \ N \ (N + 1)                  \cr
&+ {m(5) \over 362880} \ (N - 7) \ (N - 6) \ (N - 5) \ (N - 4) \ (N - 3) \ (N - 2) \ (N - 1) \ N \ (N + 1)  \ \ .  } $$
and with a straightforward computation
$$ \eqalign{
\Theta(2, \lambda_1+\lambda_2+\lambda_6, N)={3 \over g(2,N)} \ (
- & 1152 - \ 70 N + 113 \ N^2 + 4 \ N^3 + N^4)  \cr
\Theta(4, \lambda_1+\lambda_2+\lambda_6, N)={1 \over g(4,N)} \ (
- & 7925760 - 3447368 \ N - 69144 \ N^2 + 191516 \ N^3 + \cr
& 11947 \ N^4 - 2052 \ N^5 + 1154 \ N^6 + 24 \ N^7 + 3 \ N^8) \cr
\Theta(6, \lambda_1+\lambda_2+\lambda_6, N)={1 \over g(6,N)} \ (
- & 9704669184 - 9453386848 \ N - 3436715360 \ N^2 + 155802792 \ N^3 + \cr
& 289898824 \ N^4 + 6448322 \ N^5 - 10826973 \ N^6 + 375224 \ N^7 + \cr
& 259141 \ N^8 - 7110 \ N^9 + 2445 \ N^{10} + 36 \ N^{11} + 3 \ N^{12} ) \cr
\Theta(3, \lambda_1+\lambda_2+\lambda_6, N)={432 \over g(3,N)} \ (
& 448 + 98 \ N - 31 \ N^2 - 4 \ N^3 + N^4) \cr
\Theta(5, \lambda_1+\lambda_2+\lambda_6, N)={432 \over g(5,N)} \ (
& 663040 + 458420 \ N + 96638 \ N^2 - \cr
& 22556 \ N^3 - 8441 \ N^4 + 608 \ N^5 + 144 \ N^6 - 16 \ N^7 + 3 N^8) \cr
\Theta(7, \lambda_1+\lambda_2+\lambda_6, N)={288 \over g(7,N)} \ (
& 1099055104 + 1399367648 \ N + 708562192 \ N^2 + \cr
& 61441320 \ N^3 - 51346940 \ N^4 - 8957530 \ N^5 + 2041581 \ N^6 + \cr
& 268592 \ N^7 - 85565 \ N^8 - 1730 \ N^9 + 1875 \ N^{10} - 60 \ N^{11} + 9 \ N^{12} )  } $$
where
$$ g(s,N)=3 \ 2^s \ (s + 1) \ (N + 1)^{s-1} \ \ . $$
In this example, we have supressed explicit N-dependences though we recall that
all expressions are valid for $ N > 6 $. It is thus seen that all the formulas
given from (III.8) to (III.19) above give rise to the same result
$$ \eqalign{
m(1)&= \ \ 2 \ m(0)     \cr
m(2)&= \ \ 5 \ m(0)     \cr
m(3)&= \ 10 \ m(0)    \cr
m(4)&= \ 35 \ m(0)    \cr
m(5)&=105 \ m(0)   } $$
for which one always knows that $ m(0) = 1 $. As can be easily investigated
by Weyl dimension formula, this also leads us to the result
$$ dimR(\lambda_1+\lambda_2+\lambda_6) ={1 \over 3456} \ (N - 4) \ (N - 3) \
(N - 2) \ (N - 1) \ N \ (N + 1)^2 \ (N + 2) \ (N + 3) $$ .

\vskip 3mm
\noindent {\bf{V.\ CONCLUSIONS}}
\vskip 3mm
We obtained here 12 multiplicity formulas for a weight within an irreducible
representation of $A_N$ Lie algebras. The method is based essentially on some
polinomials representing eigenvalues of Casimir operators. These polinomials
has been given for degrees s=4,5,6,7 in a previous work. If one further
considers the Casimir operators of higher degrees, it is clear that one can
obtain, in essence, an infinite number of multiplicity formulas.

On the other hand, Casimir eigenvalues of any other Classical or Exceptional
Lie algebra having a subalgebra of type $A_N$ can be obtained by the aid of
these polinomials. It could therefore be expected that the multiplicity formulas
given above are to be generalized further to these Lie algebras. Another point
which seems to be worthwhile to study is the hope that such a framework will
prove useful also for Lie algebras beyond the finite ones. To this end, the
crucial point will be to fix a convenient subalgebra which underlies the
infinite dimensional one. For instance, one can think that an $E_8$
multiplicity formula could be reformulated in terms of its subalgebra $A_7$
or more suitably $A_8$. But what is more intriguing is to consider the same
possibility ,say, for hyperbolic Lie algebra $E_{10}$. It is clear that to
investigate these possibilities shed some light on the multiplicity problems
of infinite Lie algebras for which quite little is known about their
multiplicity formulas in general.

\vskip3mm
\noindent{\bf {REFERENCES}}
\vskip3mm
\noindent [1] Karadayi H.R and Gungormez M: Explicit Construction of Casimir
Operators and Eigenvalues:II ,

\noindent {\bf physics/mathematical methods in physics/9611002} , submitted to Jour.Math.Phys.

\noindent [2] Karadayi H.R and Gungormez M: Explicit Construction of Casimir
Operators and Eigenvalues:I ,

\noindent {\bf hep-th/9609060} ,submitted to Jour.Math.Phys.

\noindent [3] Kostant B. ; Trans.Am.Math.Soc. 93 (1959) 53-73

\noindent [4] Freudenthal H. ; Indag.Math. 16 (1954) 369-376 and 487-491

\noindent Freudenthal H. ; Indag.Math. 18 (1956) 511-514

\noindent [5] Patera J. and Sankoff D. : Tables of Branching Rules for
Representations of Simple Lie Algebras, L'Universite de Montreal, 1973

\noindent McKay W. and Patera J. : Tables of Dimensions, Indices and Branching Rules for
Representations of Simple Algebras, Dekker, NY 1981

\noindent Slansky R : Group Theory for Unified Model Building, Physics Reports

\noindent [6] Humphreys J.E: Introduction to Lie Algebras and Representation
Theory , Springer-Verlag (1972) N.Y.

\hfill\eject

\vskip 3mm
\noindent {\bf{ APPENDIX }}
\vskip 3mm
We give here N-dependent coefficient polinomials encountered in the multiplicity
formulas given in section (III).

$$ \eqalign{
&f^7_7(N)= N^6 + 6 \ N^5 + 50 \ N^4 + 160 \ N^3 + 309 \ N^2 + 314 \ N + 120 \cr
&f^7_{52}(N)= N^5 + 5 \ N^4 + 21 \ N^3 + 43 \ N^2 - 70 \ N - 96               \cr
&f^7_{43}(N)= N^5 + 5 \ N^4 +  9 \ N^3 +  7 \ N^2 + 62 \ N + 60               \cr
&f^7_{322}(N)= 2 \ N^4 + 8 \ N^3 - 5 \ N^2 - 26 \ N - 15  \cr
& \ \ \ \ \ \ \ \ \ \ \ \ \cr
&f^{52}_7(N)= N^7 + 7 \ N^6 + 31 \ N^5 + 85 \ N^4 + 16 \ N^3 - 236 \ N^2 - 192 \ N            \cr
&f^{52}_{52}(N)= N^8 + 8 \ N^7 + 32 \ N^6 + 80 \ N^5 + 515 \ N^4 + 1676 \ N^3 + 1648 \ N^2 + 72 \ N - 10080  \cr
&f^{52}_{43}(N)= 6 \ N^6 + 36 \ N^5 - N^2 + 13 \ N^4 - 188 \ N^3 + 470 \ N + 840                             \cr
&f^{52}_{322}(N)= N^7 + 7 \ N^6 - 70 \ N^4 + 217 \ N^3 + 987 \ N^2 - 134 \ N - 840 \cr
&f^{52}_{32}(N)= N^{10} + 10 \ N^9 - 19 \ N^8 - 392 \ N^7 - 497 \ N^6 + 3178 \ N^5 + \cr
& \ \ \ \ \ \ \ \ \ \ \ \ \ \ \ \ \ \ \ \ \ \ \ \ \ \ \ \ \ \ \ \ \ \ \ \ \ \ \
\ \ \ \ \ \ \ \ \ \ \ \ \ \ \ \ \ \ \ \
9183 \ N^4 + 6948 \ N^3 - 604 \ N^2 - 1680 \ N   \cr
&f^{52}_5(N)= N^{11} + 11 \ N^{10} - 2 \ N^9 - 348 \ N^8 - 1071 \ N^7 + 231 \ N^6 + 10856 \ N^5 + \cr
& \ \ \ \ \ \ \ \ \ \ \ \ \ \ \ \ \ \ \ \ \ \ \ \ \ \ \ \ \ \ \ \ \ \ \ \ \ \ \
\ \ \ \ \ \ \ \ \ \ \ \ \ \ \ \ \ \ \ \
35458 \ N^4 + 52712 \ N^3 + 37224 \ N^2 + 10080 \ N \cr
& \ \ \ \ \ \ \ \ \ \ \ \ \cr
&f^{43}_7(N)= N^7 + 7 \ N^6 + 19 \ N^5 + 25 \ N^4 + 76 \ N^3 + 184 \ N^2 + 120 \ N \cr
&f^{43}_{52}(N)= 6 \ N^6 + 36 \ N^5 + 13 \ N^4 - 188 \ N^3 - N^2 + 470 \ N + 840   \cr
&f^{43}_{43}(N)= N^8 + 8 \ N^7 + 16 \ N^6 - 16 \ N^5 + 681 \ N^4 + 2980 \ N^3 - 986 \ N^2 - 8060 \ N - 8400  \cr
&f^{43}_{322}(N)= 2 \ N^7 + 14 \ N^6 + 133 \ N^5 + 525 \ N^4 - 553 \ N^3 - 3647 \ N^2 + 1510 \ N + 4200 \cr
&f^{43}_3(N)= (N - 5) \ (N - 4) \ (N - 3) \ (N - 2) \ N \ (N + 1)^3 \ (N + 2) \ (N + 4)  \cr
& \ \ \ \ \ \ \ \ \ \ \ \ \ \ \ \ \ \ \ \ \ \ \ \ \ \ \ \ \ \ \ \ \ \ \ \ \ \ \
\ \ \ \ \ \ \ \ \ \ \ \ \ \ \ \ \ \ \ \ \ \ \ \ \ \ \ \ \ \ \ \ \ \ \ \ \ \ \ \
(N + 5) \ (N + 6) \ (N + 7)  \cr
& \ \ \ \ \ \ \ \ \ \ \ \ \cr
&f^{322}_7(N)= 2 \ N^6 + 12 \ N^5 + 11 \ N^4 - 36 \ N^3 - 67 \ N^2 - 30 N \cr
&f^{322}_{52}(N)= N^7 + 7 \ N^6 - 70 \ N^4 + 217 \ N^3 + 987 \ N^2 - 134 \ N - 840 \cr
&f^{322}_{43}(N)= 2 \ N^7 + 14 \ N^6 + 133 \ N^5 + 525 \ N^4 - 553 \ N^3 - 3647 \ N^2 + 1510 \ N + 4200   \cr
&f^{322}_{322}(N)= N^8 + 8 \ N^7 - 3 \ N^6 - 130 \ N^5 + 109 \ N^4 + 1452 \ N^3 + 5113 \ N^2 + 6890 \ N - 4200 \cr
&f^{322}_5(N)=(N - 5) \ (N - 4) \ N \ (N + 1)^2 \ (N + 2) \ (N + 6) \ (N + 7) \ (N^2 + 2 \ N - 1) \cr
&f^{322}_{32}(N)= - (N - 4) \ (N - 5) \ N \ (N + 1) \ (N + 2) \ (N + 6) \ (N + 7) \ (N^4 + 4 \ N^3 + 6 \ N^2 + 4 \ N + 25) \cr
&f^{322}_3(N)=(N - 5) \ (N - 4) \ N \ (N + 1) \ (N + 2) \ (N + 6) \ (N + 7) \ (5 \ N^7 + 35 \ N^6 - \cr
& \ \ \ \ \ \ \ \ \ \ \ \ \ \ \ \ \ \ \ \ \ \ \ \ \ \ \ \ \ \ \ \ \ \ \ \ \ \ \
 \ \ \ \ \ \ \ \ \ \ \ \ \ \ \ \ \ \ \ \
14 \ N^5 - 420 \ N^4 - 445 \ N^3 + 625 \ N^2 + 2014 \ N + 1320 ) \cr
& \ \ \ \ \ \ \ \ \ \ \ \ \cr
&f^6_6(N)= N^5 + 5 \ N^4 + 25 \ N^3 + 55 \ N^2 + 58 \ N + 24  \cr
&f^6_{42}(N)= N^4 + 4 \ N^3 + 7 \ N^2 + 6 \ N - 18  \cr
&f^6_{33}(N)= 3 \ N^4 + 12 \ N^3 + 7 \ N^2 - 10 \ N + 72 \cr
&f^6_{222}(N)= N^3 + 3 \ N^2 - 4 \ N - 6  } $$

$$ \eqalign{
&f^{42}_6(N)= N^7 + 7 \ N^6 + 21 \ N^5 + 35 \ N^4 + 14 \ N^3 - 42 \ N^2 - 36 \ N   \cr
&f^{42}_{42}(N)= N^8 + 8 \ N^7 + 28 \ N^6 + 56 \ N^5 + 169 \ N^4 + 452 \ N^3 +    \cr
& \ \ \ \ \ \ \ \ \ \ \ \ \ \ \ \ \ \ \ \ \ \ \ \ \ \ \ \ \ \ \ \ \ \ \ \ \ \ \
762 \ N^2 + 684 \ N - 2160   \cr
&f^{42}_{33}(N)= N^6 + 6 \ N^5 + 5 \ N^4 - 20 \ N^3 - 20 \ N^2 + 16 \ N + 96  \cr
&f^{42}_{222}(N)= 2 \ N^7 + 14 \ N^6 - 3 \ N^5 - 155 \ N^4 + 163 \ N^3 + 1221 \ N^2 - 162 \ N - 1080  \cr
&f^{42}_4(N)= N^{11} + 11 \ N^{10} + 14 \ N^9 - 204 \ N^8 - 747 \ N^7 - 189 \ N^6 + \cr
& \ \ \ \ \ \ \ \ \ \ \ \ \ \ \ \ \ \ \ \ \ \ \ \ \ \ \ \ \ \ \ \ \ \ \ \ \ \ \
3716 \ N^5 + 9334 \ N^4 + 10696 \ N^3 + 6168 \ N^2 + 1440 \ N  \cr
&f^{42}_{22}(N)= 2 \ N^{10} + 20 \ N^9 + 3 \ N^8 - 456 \ N^7 - 1008 \ N^6 + 1680 \ N^5 + \cr
& \ \ \ \ \ \ \ \ \ \ \ \ \ \ \ \ \ \ \ \ \ \ \ \ \ \ \ \ \ \ \ \ \ \ \ \ \ \ \
7327 \ N^4 + 7036 \ N^3 + 1236 \ N^2 - 720 \ N   \cr
&f^{42}_2=(N + 1)^2 \ g_6(N) \cr
& \ \ \ \ \ \ \ \ \ \ \ \ \cr
&f^{33}_6(N)= 3 \ N^7 + 21 \ N^6 + 49 \ N^5 + 35 \ N^4 + 56 \ N^3 + 196 \ N^2 + 144 \ N  \cr
&f^{33}_{42}(N)= N^6 + 6 \ N^5 + 5 \ N^4 - 20 \ N^3 - 20 \ N^2 + 16 \ N + 96  \cr
&f^{33}_{33}(N)= N^8 + 8 \ N^7 - 112 \ N^5 + 127 \ N^4 + 1404 \ N^3 + 580 \ N^2 - 2032 \ N - 3840  \cr
&f^{33}_{222}(N)= 4 \ N^5 + 20 \ N^4 - 19 \ N^3 - 137 \ N^2+ 78 \ N + 180   \cr
& \ \ \ \ \ \ \ \ \ \ \ \ \cr
&f^{222}_6(N)= N^6 + 6 \ N^5 + 7 \ N^4 - 12 \ N^3 - 26 \ N^2 - 12 \ N  \cr
&f^{222}_{42}(N)= 2 \ N^7 + 14 \ N^6 - 3 \ N^5 - 155 \ N^4 + 163 \ N^3 + 1221 \ N^2 - 162 \ N - 1080 \cr
&f^{222}_{33}(N)= 4 \ N^5 + 20 \ N^4 - 19 \ N^3 - 137 \ N^2 + 78 \ N + 180  \cr
&f^{222}_{222}(N)= N^8 + 8 \ N^7 - 7 \ N^6 - 154 \ N^5 - 79 \ N^4 + 860 \ N^3 + 1777 \ N^2 + 1338 \ N - 3240  \cr
&f^{222}_4(N)= 2 \ N^{10} + 20 \ N^9 + 3 \ N^8 - 456 \ N^7 - 1008 \ N^6 + 1680 \ N^5 + 7327 \ N^4 + \cr
& \ \ \ \ \ \ \ \ \ \ \ \ \ \ \ \ \ \ \ \ \ \ \ \ \ \ \ \ \ \ \ \ \ \ \ \ \ \ \
7036 \ N^3 + 1236 \ N^2 - 720 \ N  \cr
&f^{222}_{22}(N)= N^{11} + 11 \ N^{10} + 7 \ N^9 - 267 \ N^8 - 687 \ N^7 + 1407 \ N^6 + 5543 \ N^5 + \cr
& \ \ \ \ \ \ \ \ \ \ \ \ \ \ \ \ \ \ \ \ \ \ \ \ \ \ \ \ \ \ \ \ \ \ \ \ \ \ \
157 \ N^4 - 6664 \ N^3 + 6252 \ N^2 + 9360 \ N  \cr
&f^{222}_2(N)= 5 \ N^{14} + 70 \ N^{13} + 186 \ N^{12} - 1408 \ N^{11} - 7964 N^{10} - \cr
& \ \ \ \ \ \ \ \ \ \ \ \ \ \ \ \ \ \ \ \ \ \ \ \ \ \ \ \ \ \ \ \ \ \ \ \ \ \ \
1320 \ N^9 + 65098 \ N^8 + 121616 \ N^7 - 67617 \ N^6 - 437030 \ N^5 - 422284 \ N^4 + \cr
& \ \ \ \ \ \ \ \ \ \ \ \ \ \ \ \ \ \ \ \ \ \ \ \ \ \ \ \ \ \ \ \ \ \ \ \ \ \ \
127992 \ N^3 + 432576 \ N^2 + 190080 \ N                        \cr
& \ \ \ \ \ \ \ \ \ \ \ \ \cr
&f^5_5(N)= N^4 + 4 \ N^3 + 11 \ N^2 + 14 \ N + 6                \cr
&f^5_{32}(N)= N^3 + 3 \ N^2 + N - 1                             \cr
& \ \ \ \ \ \ \ \ \ \ \ \ \cr
&f^{32}_5(N)= N^3 + 3 \ N^2 + N - 1                             \cr
&f^{32}_32(N)= N^4 + 4 \ N^3 + 6 \ N^2 + 4 \ N + 25             \cr
&f^{32}_3(N)=(N - 3) \ (N - 2) \ (N + 1)^3 \ (N + 4) \ (N + 5)  \cr
& \ \ \ \ \ \ \ \ \ \ \ \ \cr
&f^4_4(N)= N^3 + 3 \ N^2 + 4 \ N + 2                            \cr
&f^4_{22}(N)= 2 \ N^2 + 4 \ N - 1                               \cr
& \ \ \ \ \ \ \ \ \ \ \ \ \cr
&f^{22}_4(N)= 2 \ N^3 + 6 \ N^2 + 3 \ N - 1                              \cr
&f^{22}_{22}(N)= N^4 + 4 \ N^3 - 8 \ N + 13                              \cr
&f^{22}_2(N)= N^7 + 7 \ N^6 + 8 \ N^5 - 30 \ N^4 - 59 \ N^3 - N^2 + 50 \ N + 24  }   $$

\end